\documentclass[reprint,aps,prb,nofootinbib,showkeys,showpacs]{revtex4-1}
\usepackage{scrextend}

\usepackage{amssymb}                                    
\usepackage{fullpage}
\usepackage{graphicx}  
\usepackage{caption}                                        
\usepackage[squaren,Gray, cdot]{SIunits}        
\usepackage{amsmath}
\usepackage{natbib}
\usepackage{bm}
\usepackage{hyperref}

\begin{document}

\title{Density-based crystal plasticity : from the discrete to the continuum}                       

\author{Pierre-Louis Valdenaire}
\author{Yann Le Bouar}
\author{Beno\^it Appolaire}
\author{Alphonse Finel}
        \email{alphonse.finel@onera.fr}
\affiliation{Laboratoire d'Etude des Microstructures, ONERA/CNRS, 29, avenue de la division Leclerc, 92322 Ch\^{a}tillon, France}  

\date{\today}   

\begin{abstract}
Because of the enormous range of time and space scales involved in dislocation dynamics, plastic modeling at macroscale requires a continuous formulation. In this paper, we present a rigorous formulation of the transition between the discrete, where plastic flow is resolved at the scale of individual dislocations, and the continuum, where dislocations are represented by densities. First, we focus on the underlying coarse-graining procedure and show that the emerging correlation-induced stresses are scale-dependent.  Each of these stresses can be expanded into the sum of two components. The first one depends on the local values of the dislocation densities and always opposes the sum of the applied stress and long-range mean field stress generated by the geometrically necessary dislocation (GND) density; this stress acts as a friction stress. The second component depends on the local gradients of the dislocation densities and is inherently associated to a translation of the elastic domain; therefore, it acts as a back-stress. We also show that these friction and back-  stresses contain symmetry-breaking components that make the local stress experienced by dislocations to depend on the sign of their Burgers vector. 

\end{abstract}




\maketitle

\section{Introduction}
Plasticity of crystalline solids involves the notion of dislocations. However, even today, conventional plasticity theories use mesoscopic variables and evolution equations that do not involve dislocations. This paradoxical situation is due to the enormous length and time scales that separate the description of plasticity at the level of individual dislocations and the macroscopic scale of engineering materials. This huge space and time separation renders the hope to use a discrete dislocation based approach out of reach for treating engineering problems. It could be argued that conventional or phenomenological plasticity theories are justified because, at the macroscopic scale, engineering materials always display some sort of disorder that gives to any macroscopic property or measure an inevitable averaging character. Hence, at macroscale, plastic strain may be seen as resulting from a space and time average over a huge number of individual dislocation glide events.

Nevertheless, conventional plasticity theories rely on strong approximations and on phenomenological laws that must be calibrated for each material or for each specific applications. Therefore, it is desirable to make a link between the micro and macro scales and to develop a mesoscopic plasticity theory that relies on a sound physical basis, i.e. that at least incorporates dislocation glide. The development of such a mesoscale theory is also crucial to better understand and simulate the materials behavior at length scales where the elastic interaction between dislocations becomes of the order of the interaction between dislocations and obstacles, such as precipitates in a matrix, small grains in a polycrystal or interfaces in nano-materials. At these scales, dislocations display collective phenomena that result in patterning and complex dynamic regimes. In these situations, plasticity cannot be described by a simple averaged plastic strain that obeys local time-dependent equations. Size-dependent effects and, most importantly, transport become fondamental. Conventional theories of plasticity are no longer valid and are unable to account for the complexity of the plastic activity because they lack the relevant internal length scale and do not incorporate transport.

These considerations motivate the development of continuum models in which dislocations are represented by continuous densities and in which the dynamics has conserved the transport character of the underlying dislocation glide.

Continuum dislocation representations often start from the Nye \cite{NY1953} and Kr\"oner \cite{KR1959} representation of dislocations. This is the case of the Field Dislocation Model (FDM) proposed by Acharya \cite{AC2001,AC2003} and developed subsequently by various authors \cite{RO2005,FR2011,TAU2013,FR2014}. The basic equations have been in fact known as early as the $60$'s \cite{KO1962,MU1963} (see also Ref.~\onlinecite{LA1970,NA1979}). The basic ingredient of the FDM is the dislocation density tensor $\alpha\!=\!-\operatorname{curl}\beta^p$, where $\beta^p$ is the plastic distorsion tensor. When envisaged at the smallest scale, the tensor $\alpha$ represents all the dislocations and there is no need to introduce the concept of "geometrically necessary" or "statistically stored" dislocations (GND and SSD, respectively). The model is then exact, regardless of the atomic nature of the dislocations and provided that we accept that the dislocation velocity is simply proportional to the local resolved shear stress. However, being continuous by nature, the implementation of the model requires the use of a computational grid with a grid step significantly smaller than the Burgers vector length. This drastically limits the spatial length scale that can be investigated. Therefore, in order to reach a convenient macro scale, a change of scale must be performed to bridge the gap between the singular density tensor introduced above and a continuous one defined at an intermediate scale. There is of course no unique way to select this so-called "mesoscale". Obviously, the mesoscale must be larger than the average distance between dislocations and smaller than the characteristic length scale we want to investigate (average grain size in polycrystals, average distance between interfaces in multiphase alloys, etc.). The underlying averaging or "coarse-graining" procedure has of course been already mentioned in the context of the FDM \cite{AC2006,FR2010}.

The crucial point is that the application of the coarse-graining procedure to the FDM equations leads to transport equations for the averaged one-body GND density in which the plastic strain rate inevitably depends on the correlations between the lower scale GND and velocity fields. This closure problem is often resolved by using a phenomenological velocity law borrowed from macroscopic plasticity models leading to the so-called Phenomenological Mesoscopic Field Dislocation Model (PMFDM) \cite{AC2006,RO2006}. The actual implementation of the mesoscale FDM thus suffers from the lack of a mathematically justified mesoscale plastic strain rate. 

A more recent formulation of a Continuum Dislocation Dynamics (CDD) has been proposed by Hochrainer and its collaborators \cite{HO2006,HO2007}. It is based on a modified definition of the dislocation density tensor, in order to keep at mesoscale information concerning the geometry of the dislocations (in particular, line directions and curvatures). The necessity of using an averaging procedure to obtain a meaningful continuum model has also be pointed out in the context of the CDD formulation \cite{SA2011} (see also \cite{AZ2000,AZ2006}), but a rigorous mathematical formulation of this coarse-graining procedure has not yet been proposed.

The first attempt to better treat the closure problem has been proposed by Groma  and its collaborators \cite{1GR1997,ZA2001,2GR2002}. This is the route that we follow below. A particular attention will be paid to the nature of the coarse-graining procedure and its consequences on the local stress fields that emerge from the averaging process. We show that the emerging local friction and back-stresses, which are reminiscent of the dislocation-dislocation correlations, depend on the length scale associated to the averaging process required by the coarse-graining procedure. We also show that these correlation-induced stresses contain symmetry-breaking components that make the local stress experienced by dislocations to depend on the sign of their Burgers vector. Finally, we find that the emerging back-stress depends on the gradients of both the geometrically necessary and total dislocation densities. A brief version of these results has been presented in Ref.~\onlinecite{FI2014}.

\section{Mesoscale density-based theory \label{microdefdeeqdedis}}

We first clarify the mathematics and physical aspects of the coarse-graining procedure that must be used to coarse-grain the dislocation dynamics from the discrete to the continuum. We consider the simplest situation, namely a 2D dislocation system with $N$ edge dislocation lines parallel to the $z$-axis restricted to glide along the $x$-axis. The Burgers vector of dislocation $i$, $i\!=\!1$ to $N$, is noted $s_i\vec b$, where $s_i\!$ is the sign of the dislocation $i$ and $\vec b\!=\!(b,0,0)$. We assume an overdamped motion: the glide velocity of the $i^{th}$ dislocation along the $x$-axis is simply proportional to the resolved Peach-Koehler force acting on the dislocation $i$,
\begin{equation}
\label{eq:vitessedislodetail}
\frac{d\vec r_i}{dt}=Ms_i\vec b\left(\sum_{j\neq i}^Ns_j\tau_{ind}(\vec r_i-\vec r_j)+\tau_{ext}\right),
\end{equation}
where $M$ is the mobility coefficient equal to the inverse of the dislocation drag coefficient, $\tau_{ext}$ the external stress resolved in the slip system and $\tau_{ind}(\vec r)$ the resolved shear stress at position $\vec r$ generated by a positive dislocation located at the origin:
\begin{equation}
\label{eq:defAFdetauind}
\tau_{ind}(x,y)=\frac{\mu b}{2\pi(1-\nu)}\frac{x(x^2-y^2)}{(x^2+y^2)^2}
\end{equation}
where $\mu$ is the shear modulus and $\nu$ the Poisson ratio.

The first step is to define discrete dislocation densities:
\begin{equation}
\label{eq:AFrhodisc}
\begin{matrix}
\rho^+_{dis}(\vec r,t,\{\vec r^{\,0}_k\})=\sum_{i=1}^N\delta_{s_i,+1}\delta(\vec r-\vec r_i(t,\{\vec r^{\,0}_k\}))
\\
\rho^-_{dis}(\vec r,t,\{\vec r^{\,0}_k\})=\sum_{i=1}^N\delta_{s_i,-1}\delta(\vec r-\vec r_i(t,\{\vec r^{\,0}_k\}))
\end{matrix}
\end{equation}
where $\{\vec r^{\,0}_k\}$ refers to the initial positions of the $N$ dislocations, $\delta_{s,t}$ is the Kronecker symbol and $\delta(\vec r)$ the 2D Dirac function. The notation $\vec r_i(t,\{\vec r^{\,0}_k\})$ means that the trajectory of dislocation $i$ depends on the initial dislocation positions $\{\vec r^{\,0}_k\}$.

By multiplying Eq.~(\ref{eq:vitessedislodetail}) by the Dirac function $\delta(\vec r-\vec r_i(t,\{\vec r^{\,0}_k\}))$ and taking its derivative with respect to $\vec r$, we get the following transport equation for the discrete densities:
\begin{widetext}
\begin{equation}
\label{eq:AFeqrhodisc}
-\frac{\partial}{\partial t}\rho^s_{dis}(\vec r)=sM\vec b\cdot \frac{\partial}{\partial\vec r}\left\{\int\limits_{\vec r\,'\neq\vec r}\tau_{ind}(\vec r-\vec r\,')\sum_{s'=\pm1}s'\rho_{dis}^{s'}(\vec r\,')\rho_{dis}^{s}(\vec r)d\vec r\,'+\tau_{ext}\,\rho^s_{dis}(\vec r)\right\}
\end{equation}
\end{widetext}
where, to simplify the notation, we write $\rho^s_{dis}(\vec r)$ for $\rho^s_{dis}(\vec r,t,\{\vec r^{\,0}_k\})$. Obviously, these transport equations link the time-dependence of the one-body densities to the products of two one-body densities, which is a direct consequence of the pairwise dislocation interactions. At this stage, the dislocation densities $\rho^s_{dis}(\vec r)$ are highly singular. The next step is to introduce a coarse-graining procedure.

\subsection{Coarse-graining procedure\label{subsectionCG}}

We introduce now a coarse-graining procedure commonly used in statistical physics (see, for exemple, Ref.~\onlinecite{MA2003}). We first define a space and time convolution process that we use to coarse-grain microscopic fields to mesoscopic ones:
\begin{equation}
\label{eq:AFcg}
f_{meso}(\vec r,t)=\iint w(\vec r\,',t')\ f_{micro}(\vec r+\vec r\,',t+t')\ d\vec r\,'dt'.
\end{equation}
where the weighting function $w(\vec r,t)$ is normalized. For simplicity, and without loss of generality, we choose $w(\vec r,t)$ to be separable:
\begin{equation}
\label{eq:AFfuncw}
w(\vec r,t)=w_L(\vec r)\ w_{T(L)}(t)
\end{equation}
where the functions $w_L(\vec r)$ and $w_{T(L)}(t)$ are separately normalized:
\begin{equation}
\label{eq:AFnormdefuncw}
\int w_L(\vec r)\ d\vec r=1
\text{ and }
\int w_{T(L)}(t)\ dt=1.
\end{equation}
The spatial linear dimension $L$ of $w_L(\vec r)$ should be of the order of the spatial resolution of the continuous model we seek and, obviously, significantly larger than the average distance between dislocations. The temporal width $T(L)$ of the time window $w_{T(L)}(t)$ should, in all generality, depend on $L$. In fact, the appropriate choice of $T(L)$ is linked to the kinetic behaviour of the degrees of freedom that, inevitably, we will have to average out in order to close the theory:  $T(L)$ should be defined in such a way that the correlations we want to average out have the time to reach a stationary state at scale $L$.  We comment on that point in section~\ref{correldeAF}. Here, we just mention that, for convenience, we choose $w_{T(L)}(t)$ to be non-zero only for $t\!\leqslant\!0$:
\begin{equation}
\label{eq:AFdefspedewtl}
w_{T(L)}(t)\neq0\quad\text{if}\quad t\leqslant0.
\end{equation}

Mesoscopic density fields may be defined through Eq.~(\ref{eq:AFcg}), but this is not enough to get a consistent continuous transport theory. First, we expect that the time evolution of the mesoscopic dislocation densities will be given by first-order transport (i.e. hyperbolic) equations. These equations must be supplemented by initial conditions at $t=0$ which, of course, must be defined at mesoscale. In other words, the coarse-graining procedure should be such that, when applied to Eq.~(\ref{eq:AFeqrhodisc}) and its initial condition given by the dislocation positions $\{\vec r^{\,0}_k\}$ at $t\!=\!0$, we end up with a set of mesoscopic transport equations supplemented by continuous initial conditions that do not depend on any specific initial set $\{\vec r^{\,0}_k\}$. Therefore, if $\rho^s(\vec r,t\!=\!0)$, $s\!=\!\pm1$, are given initial continuous densities, we must introduce a $N$-body probability density distribution $P_{\{s^{\,0}_k\}}(\vec r^{\,0}_1,\dots,\vec r^{\,0}_N)$ on the (discrete) initial positions $\{\vec r^{\,0}_k\}$ which is linked to the initial mesoscopic densities $\rho^s(\vec r,t\!=\!0)$ in a way that we discuss below. The distribution $P_{\{s^{\,0}_k\}}(\vec r^{\,0}_1,\dots,\vec r^{\,0}_N)$, where $\{s^{\,0}_k\}$ refers to the predefined (and fixed) signs of the $N$ dislocations, introduces a statistical ensemble on the initial discrete dislocation positions: $P_{\{s^{\,0}_k\}}(\vec r^{\,0}_1,\dots,\vec r^{\,0}_N)d\vec r^{\,0}_1\dots d\vec r^{\,0}_N$ is the probability to have an initial dislocation configuration with dislocation $1$, whose sign is $s_1$, in a small volume $d\vec r^{\,0}_1$ around position $\vec r^{\,0}_1$, dislocation $2$, whose sign is $s_2$,  in a small volume $d\vec r^{\,0}_2$ around position $\vec r^{\,0}_2$, etc.

Now, the overall coarse-graining procedure is defined as the conjugate action of the space-time convolution window $w(\vec r,t)$ and the ensemble average defined by the probability density $P_{\{s^{\,0}_k\}}(\vec r^{\,0}_1,\dots,\vec r^{\,0}_N)$. The mesoscopic field $X_{meso}(\vec r,t)$ associated with the discrete field $X_{dis}(\vec r,t,\{\vec r^{\,0}_k\})$ is therefore defined by:
\begin{widetext}
\begin{equation}
\label{eq:AFdefdefuncmeso}
X_{meso}(\vec r,t)=\prod_{k=1}^N\int d\vec r^{\,0}_kP_{\{s^{\,0}_k\}}(\vec r^{\,0}_1,\dots,\vec r^{\,0}_N)\int d\vec r\,'\int dt'w(\vec r\,',t')X_{dis}(\vec r+\vec r\,',t+t',\{\vec r^{\,0}_k\}).
\end{equation}
\end{widetext}
We refer to this coarse-graining procedure by the following short-hand notation:
\begin{equation}
\label{eq:AFdefdefuncmeso2}
X_{meso}(\vec r,t)=\langle\langle X_{dis}(\vec r,t,\{\vec r^{\,0}_k\})\rangle\rangle_P
\end{equation}
where the double brakets refer to the space and time convolution and the lower index $P$ to the ensemble average. The mesoscopic one-body and two-body densities are therefore defined by:
\begin{equation}
\label{eq:AFdefderhoun}
\rho^s(\vec r,t)=\langle\langle\rho^s_{dis}(\vec r,t,\{\vec r^{\,0}_k\})\rangle\rangle_P
\end{equation}
and
\begin{equation}
\label{eq:AFdefderhodeux}
\rho^{ss'}\!(\vec r,\vec r\,',t)=\langle\langle\rho^{s}_{dis}(\vec r,t,\{\vec r^{\,0}_k\})\rho^{s'}_{dis}(\vec r\,',t,\{\vec r^{\,0}_k\})\rangle\rangle_P.
\end{equation}
We mention that the two-body densities defined in Eq.~(\ref{eq:AFdefderhodeux}) are continuous function of $\vec r$ and $\vec r\,'$. This would not be the case if the coarse-graining procedure was limited to a space and time convolution. This is the second reason why we need to consider an average over a statistical ensemble.

We can now precise the link, mentioned above, between the probability density $P_{\{s^{\,0}_k\}}(\vec r^{\,0}_1,\dots,\vec r^{\,0}_N)$, that defines the statistical ensemble, and the continuous dislocation densities $\rho^s(\vec r,t)$ that will be used as initial conditions for the mesoscopic kinetic equations. We consider that any discrete initial condition $\{\vec r^{\,0}_k\}$ on the $N$ dislocation positions is extended to $t\!<\!0$:
\begin{equation}
\label{eq:AFcondit}
i=1\text{ to }N\quad\text{and}\quad t\leqslant0:\ \vec r_i(t,\{\vec r^{\,0}_k\})=\vec r^{\,0}_i.
\end{equation}
Then, using the definition of the discrete densities (Eq.~(\ref{eq:AFrhodisc})) and the definition of the coarse-grained ones (Eq.~(\ref{eq:AFdefderhoun})), we get:
\begin{widetext}
\begin{equation}
\label{eq:AFsuivant}
\rho^s(\vec r,t=0)=\prod_{k=1}^N\int d\vec r^{\,0}_kP_{\{s^{\,0}_k\}}(\vec r^{\,0}_1,\dots,\vec r^{\,0}_N)\int d\vec r\,'\int dt'w(\vec r\,',t')\sum_{i=1}^N\delta_{s_i,s}\delta(\vec r+\vec r\,'-\vec r_i(t',\{\vec r^{\,0}_k\})).
\end{equation}
\end{widetext}
Using Eqs (\ref{eq:AFfuncw}), (\ref{eq:AFnormdefuncw}) and (\ref{eq:AFcondit}), we obtain:
\begin{widetext}
\begin{equation}
\label{eq:AFsuivant2}
\rho^s(\vec r,t=0)=\sum_{i=1}^N\delta_{s_i,s}\prod_{k=1}^N\int d\vec r^{\,0}_k\ P_{\{s^{\,0}_k\}}(\vec r^{\,0}_1,\dots,\vec r^{\,0}_N)\ w_L(\vec r^{\,0}_i-\vec r).
\end{equation}
\end{widetext}
This equation constitutes a constraint that $P_{\{s^{\,0}_k\}}(\vec r^{\,0}_1,\dots,\vec r^{\,0}_N)$ must fulfill for a given set of initial mesoscopic densities $\rho^s(\vec r,t\!=\!0)$. However, this  is not enough to completely define the probability density $P$. In order to proceed, supplemental properties must be assigned to $P$. As in Ref.~\onlinecite{ZA2015}, we argue that, in order to use no more information than the one actually embedded into the mesoscopic initial densities, which in principle are meant to reflect a realistic experimental situation, the supplemental rule needed to completely define $P$ should simply invoke the maximum entropy principle. This is equivalent to impose that no other information, besides that given by the constraint of Eq.~(\ref{eq:AFsuivant2}), should be used to define the statistical ensemble associated to $P$. This implies that the stochastic variables $\vec r^{\,0}_i$, $i\!=\!1$ to $N$, must be considered as statistically independent. Therefore, they must follow one-body distribution functions $f_{s_i}(\vec r)$, that depend only on their sign $s_i$, over which the density $P_{\{s^{\,0}_k\}}(\vec r^{\,0}_1,\dots,\vec r^{\,0}_N)$ is factorized:
\begin{equation}
\label{eq:AFsuivant3}
P_{\{s^{\,0}_k\}}(\vec r^{\,0}_1,\dots,\vec r^{\,0}_N)=f_{s_1}(\vec r^{\,0}_1)f_{s_2}(\vec r^{\,0}_2)\dots f_{s_N}(\vec r^{\,0}_N).
\end{equation}
Of course, the distribution functions $f_s(\vec r)$, $s=\pm 1$,  are separately normalized:
\begin{equation}
\label{eq:AFnomrdehsad}
\int f_s(\vec r)d\vec r=1.
\end{equation}
Using Eqs.~(\ref{eq:AFsuivant3}) and (\ref{eq:AFnomrdehsad}), Eq.~(\ref{eq:AFsuivant2}) becomes
\begin{equation}
\label{eq:AFbecome}
\rho^s(\vec r,t=0)=\sum_{i=1}^N\delta_{s_i,s}\int w_L(\vec r^{\,0}_i-\vec r)f_{s_i}(\vec r^{\,0}_i)d\vec r^{\,0}_i
\end{equation}
which may be written as
\begin{equation}
\label{eq:AFbecome2}
\rho^s(\vec r,t=0)=N^s\int w_L(\vec r_0-\vec r)f_s(\vec r_0)d\vec r_0
\end{equation}
where $N^s$ is the number of dislocations of sign $s$. Up to the coefficient $N^s$, the initial condition $\rho^s(\vec r,t\!=\!0)$ is simply equal to the convolution of $f_s(\vec r)$, the distribution of initial positions of the discrete dislocations of sign $s$, with the convolution window $w_L(\vec r)$. For given set of initial conditions $\rho^s(\vec r, t=0)$, $s=\pm 1$, and a given convolution window $w_L$, Eq.~(\ref{eq:AFbecome2}) defines a unique set of functions $f_s(\vec r)$, $s=\pm 1$ and, therefore, a unique probability density $P_{\{s^{\,0}_k\}}(\vec r^{\,0}_1,\dots,\vec r^{\,0}_N)$. Thus, for prescribed initial mesoscopic dislocation densities $\rho^s(\vec r,t\!=\!0)$ and a given spatial convolution window $w_L(\vec r)$, the coarse-graining procedure introduced in Eq.~(\ref{eq:AFdefdefuncmeso}) is completely and uniquely defined.

\subsection{Coarse-grained kinetic equations\label{para142}}
By a direct application to Eq.~(\ref{eq:AFeqrhodisc}) of the coarse-graining procedure defined in Eq.~(\ref{eq:AFdefdefuncmeso}), we get the following mesoscopic equations:
\begin{widetext}
\begin{equation}
\label{eq:AFmesoeq}
-\frac{\partial}{\partial t}\rho^s(\vec r,t)=sM\vec b \cdot \frac{\partial}{\partial \vec r}\left\{\int\limits_{\vec r\,'\neq\vec r}\tau_{ind}(\vec r-\vec r\,')\sum_{s'}s'\rho^{ss'}\!(\vec r,\vec r\,',t)d\vec r\,'+\tau_{ext}\ \rho^s(\vec r,t)\right\}
\end{equation}
\end{widetext}
where the mesoscopic one-body and two-body densities $\rho^s(\vec r,t)$ and $\rho^{ss'}(\vec r,\vec r\,',t)$ have been defined in Eq.~(\ref{eq:AFdefderhoun}) and (\ref{eq:AFdefderhodeux}).\\
At this stage, no approximation has been introduced. Eq.~(\ref{eq:AFmesoeq}) is exact and contains the same information and complexity as Eq.~(\ref{eq:AFeqrhodisc}) and, therefore, as Eq.~(\ref{eq:vitessedislodetail}). However, the time evolution of one-body densities $\rho^s(\vec r,t)$ is linked to the two-body dislocation densities $\rho^{ss'}\!(\vec r,\vec r\,',t)$. It is straightforward to realize that the time evolution of these two-body densities are themselves linked to the three-body densities, and so forth. Obviously, we are faced by the classical problem of closure that we meet in statistical physics when we try to replace a set of discrete degrees of freedom by a set of continuous densities.

The next step is to solve the closure problem. This of course requires the introduction of some approximations. One way to do that is to analyse and possibly approximate the two-body correlations, defined by:
\begin{equation}
\label{eq:defdescorre}
d^{ss'}\!(\vec{r},\vec{r}\,',t)=\frac{\rho^{ss'}\!(\vec{r},\vec{r}\,',t)}{\rho^s(\vec{r},t)\ \rho^{s'}(\vec{r}\,',t)}-1.
\end{equation}
Using Eq.~(\ref{eq:defdescorre}), the kinetic equation (\ref{eq:AFmesoeq}) becomes:
\begin{widetext}
\begin{equation}
\label{eq:eqdensiteaveccorrefactorise}
-\frac{\partial}{\partial t}\rho^s(\vec{r},t)=sM\vec b\cdot\frac{\partial}{\partial\vec{r}}\Bigg[\rho^s(\vec{r},t)\left\{\tau_{ext}+\tau_{sc}(\vec{r},t)+\tau^{\,s}_{corr}(\vec{r},t)\right\}\Bigg]\!.
\end{equation}
\end{widetext}
where the local stresses $\tau^{\,s}_{sc}(\vec{r},t)$ and $\tau^{\,s}_{corr}(\vec{r},t)$ are defined by:
\begin{equation}
\label{eq:deftausc}
\tau_{sc}(\vec{r},t)=\sum_{s'}s'\int\limits_{\vec{r}\,'\neq\vec{r}}\tau_{ind}(\vec{r}-\vec{r}\,')\ \rho^{s'}(\vec{r}\,',t)\ d\vec{r}\,'
\end{equation}
and
\begin{eqnarray}
\label{eq:corredetaildependence}
\tau^{\,s}_{corr}(\vec{r},t)&=&\sum_{s'}s'\int\limits_{\vec{r}\,'\neq\vec{r}}\tau_{ind}(\vec{r}-\vec{r}\,')\ d^{ss'}\!(\vec{r},\vec{r}\,',t)\   \nonumber  \\
&\times& \rho^{s'}(\vec{r}\,',t)\ d\vec{r}\,'.
\end{eqnarray}

\subsection{Mean field stress\label{meanfieldapprox}}
Together with Eqs.~(\ref{eq:defdescorre}), (\ref{eq:deftausc}) and (\ref{eq:corredetaildependence}), kinetic equation (\ref{eq:eqdensiteaveccorrefactorise}) is exact but not closed. The simplest way to have a closed continuous theory is to neglect the correlations $d^{ss'}\!(\vec r,\vec r\,',t)$. Eqs.~(\ref{eq:eqdensiteaveccorrefactorise}) become:
\begin{equation}
\label{eq:AFnouvelleeqmf}
-\frac{\partial}{\partial t}\rho^s(\vec{r},t)=sM\vec b \cdot\frac{\partial}{\partial\vec{r}}(\rho^s(\vec{r},t)\left\{\tau_{sc}(\vec{r},t)+\tau_{ext}\right\}).
\end{equation}
The local stress exerted on dislocations of sign $s$ does not depend on $s$ and is simply the sum of the external stress $\tau_{ext}$ and the stress $\tau_{sc}(\vec r,t)$ generated by all the one-body densities and defined in Eq.~(\ref{eq:deftausc}):
\begin{equation}
\label{eq:AFnouveldeftausc}
\tau_{sc}(\vec r,t)=\int\limits_{\vec r\,'\neq\vec r}\tau_{ind}(\vec r-\vec r\,')\kappa(\vec r\,',t)d\vec r\,'
\end{equation}
where we have introduced the polar or GND (Geometrically Necessary Dislocation) density:
\begin{equation}
\label{eq:AFdefdekapp}
\kappa(\vec r,t)=\sum_{s'}s'\rho^{s'}(\vec r,t).
\end{equation}
As $\tau_{sc}(\vec r,t)$ does not incorporate any correlation effects, it may be called a mean field stress or, as it closes the theory, a self-consistent stress \cite{1GR1997}.

\subsection{Correlation-induced local stresses\label{correldeAF}}

We want now to go beyond the mean field approximation and incorporate the correlations. In other words, the correlation stress $\tau^s_{corr}(\vec r,t)$ defined in Eq.~(\ref{eq:corredetaildependence}) is now taken into account. These correlations should be approximated in order to close the theory.

For that purpose, we need to discuss the time and spatial variations of the correlation functions $d^{ss'}\!(\vec r,\vec r\,',t)$. It has already been observed \cite{ZA2001,2GR2002} that the correlation length of $d^{ss'}\!(\vec r,\vec r\,',t)$ is finite and of the order of a few average dislocation spacings. Consequently, if the width of the convolution window is sufficiently larger than the mean dislocation spacing, the correlations $d^{ss'}\!(\vec r,\vec r\,',t)$, for a fixed point $\vec r$ and as a function of $\vec r\,'$, decrease to zero before the one-body densities $\rho^{s'}(\vec r\,')$ vary significantly. Therefore, within the domain around point $\vec r$ where they are non-zero, $d^{ss'}\!(\vec r,\vec r\,',t)$ may be considered as a function of $(\vec r\!-\!\vec r\,')$ and of the local one-body densities $\rho^s(\vec r,t)$:
\begin{equation}
\label{eq:defdeffefAF}
d^{ss'}\!(\vec r,\vec r\,',t)\simeq d^{ss'}\!(\vec r-\vec r\,',\{\rho^s(\vec r,t)\},t)
\end{equation}
where the notation $\{\rho^s(\vec r,t)\}$ refers to $\{\rho^s(\vec r,t),s\!=\!\pm1\}$. Now, we comment on the time dependence of the correlations. We recall that the coarse-graining procedure introduced above (see Eqs.~(\ref{eq:AFfuncw}) and (\ref{eq:AFdefdefuncmeso})) involves a time convolution. A width $T(L)$ for the time window must be selected. 

Our present purpose is to close the theory at the order of the two-body correlations. In other words, we want to incorporate two-body correlations in such a way that their time dependence is formally linked to the time dependence of the one-body densities, which themselves are defined at scale $L$. Therefore, the time convolution should be such that the averaging process incorporates all the time scales associated to the kinetics up to spatial scale $L$. This is essential for capturing and embedding properly the lower scale kinematics and configurational dislocation properties into a physically sensitive theory where the correlations are expressed as \emph{local} functionals of one-body dislocation densities defined at scale $L$. In physical terms, this requires to select a time window $T(L)$ such that the coarse-grained correlations reach a steady state at scale $L$. 

This point should be analysed in light of the very complex spatio-temporal behaviour that dislocations often display. Their dynamics is in particular characterised by the existence of a yielding transition when they are subject to an increasing stress. Both below the yielding point and in the subsequent flowing regime, the collective dislocation motion exhibits strongly intermittent avalanche-like dynamics characterised by a slow relaxation process. It has been in particular observed  \cite{IS2011,IS2014} that, close to the yielding point but also far below, the dynamics is characterised by power laws and, therefore, is essentially scale-free up to a cut-off time $t_c(L)$ that depends essentially on the system size $L$. This size-dependent relaxation time marks a cross-over from a regime where the strain rate follows a power law, $\dot \gamma(t) \sim t^{-2/3}$, to a regime where the strain rate decays exponentially to zero or reaches a steady value, depending on whether the stress is below or above the yielding point. Therefore, a convenient choice for the time convolution window is to select a width $T(L)$ of the order of the relaxation time $t_c(L)$. Under this condition, the overall coarse-graining procedure will generate correlations which are dependent on the local one-body densities only: the explicit time dependence in $d^{ss'}(\vec r,\vec r\,',t)$ disappears and shows up only implicitly through the time dependence of the one-body densities $\rho^s(\vec r,t)$. In short, Eq.~(\ref{eq:defdeffefAF}) becomes:
\begin{equation}
\label{eq:defdeffefAF2}
d^{ss'}\!(\vec r,\vec r\,',t)\simeq d^{ss'}\!(\vec r-\vec r\,',\{\rho^s(\vec r,t)\}).
\end{equation}

Now, using again the short-range nature of the correlations discussed above, we note that $\rho^{s'}(\vec r\,',t)$ in Eq.~(\ref{eq:corredetaildependence}) may be expanded to $1^{\text{st}}$-order around $\vec r$. The local stress defined in Eq.~(\ref{eq:corredetaildependence}) is then split into two terms:
\begin{equation}
\label{eq:decompodetcorreuh}
\tau^{\,s}_{corr}(\vec{r},t)=-\tau^{\,s}_b(\vec{r},t)-\tau^{\,s}_f(\vec{r},t)
\end{equation}
with
\begin{widetext}
\begin{equation}
\label{eq:AFdeftauf}
\tau^{\,s}_f(\vec{r},t)=-\sum_{s'}s'\rho^{s'}(\vec r,t)  \int\limits_{\vec r\,'\neq\vec r}\tau_{ind}(\vec r-\vec r\,')\ d^{ss'}\!(\vec r-\vec r\,',\{\rho^s(\vec r,t)\})\ d\vec r\,'
\end{equation}
and
\begin{equation}
\label{eq:AFdeftaub}
\tau^{\,s}_b(\vec{r},t)=-\sum_{s'}s'\frac{\partial\rho^{s'}(\vec r,t)}{\partial\vec r} \cdot \int\limits_{\vec r\,'\neq\vec r}(\vec r\,'-\vec r)\ \tau_{ind}(\vec r-\vec r\,')\ d^{ss'}\!(\vec r-\vec r\,',\{\rho^s(\vec r,t)\})\ d\vec r\,'.
\end{equation}
At this stage, the coarse-grained kinetic equation given in Eq.~(\ref{eq:eqdensiteaveccorrefactorise}) reads:
\begin{equation}
\label{eq:AFneweqq}
-\frac{\partial}{\partial t}\rho^s(\vec r,t)=sM\vec b  \cdot \frac{\partial}{\partial\vec r}\left[\rho^s(\vec r,t)\left\{\tau_{ext}+\tau_{sc}(\vec r,t)-\tau_f^s(\vec r,t)-\tau^s_b(\vec r,t)\right\}\right]
\end{equation}
\end{widetext}
where the local stresses $\tau_{sc}(\vec r,t)$, $\tau_f^s(\vec r,t)$ and $\tau^s_b(\vec r,t)$ are defined in Eqs.~(\ref{eq:deftausc}), (\ref{eq:AFdeftauf}) and (\ref{eq:AFdeftaub}). Next, we discuss the physical meaning of the correlation-induced stresses $\tau^s_f$ and $\tau^s_b$.

\subsection{Physical meaning of the correlation-induced stresses $\tau^s_f$ and $\tau_b^s$\label{taufs_tau_bs}}

The physical meaning and properties of these local stresses will of course be inherited from the symmetry properties of the correlations. It should also be clear that these correlations depend on the stress experienced by the dislocations.  Within the spirit of the present coarse-graining procedure, which inevitably leads to a hierarchy of independent and successive many-body densities, we consider that the stress dependence of the $k$-body densities is due to the stress generated by the correlations up to order $(k\!-\!1)$. Therefore, the stress dependence of the correlations $d^{ss'}$ is due to the sum of the external stress and the mean-field stress $\tau_{sc}(\vec r,t)$. We note $\tau_{lo}(\vec r,t)$ this \emph{low-order} stress: $\tau_{lo}(\vec r,t)\!=\!\tau_{ext}\!+\!\tau_{sc}(\vec r,t)$.

Using the discrete kinetic equation (\ref{eq:vitessedislodetail}) and its symmetry properties, it is easy to show that  the correlations display the following property:
\begin{widetext}
\begin{equation}
\label{eq:defsym3AF}
d^{ss'}\!(x-x',y-y',\{\rho^s(\vec r,t)\},\tau_{lo}(\vec{r},t))=d^{ss'}\!(x'-x,y-y',\{\rho^s(\vec r,t)\},-\tau_{lo}(\vec{r},t))
\end{equation}
\end{widetext}
where the dependence of the correlations on the low-order stress  $\tau_{lo}(\vec r, t)$ has been explicitly pointed out. Also, according to their very definition (Eq.~\ref{eq:AFdefderhodeux}), we obviously have:
\begin{widetext}
\begin{equation}
\label{eq:defsym4AF}
d^{ss'}\!(x-x',y-y',\{\rho^s(\vec r,t)\},\tau_{lo}(\vec{r},t))=d^{s'\!s}(x'-x,y'-y,\{\rho^s(\vec r,t)\},\tau_{lo}(\vec{r},t)).
\end{equation}
\end{widetext}
For later use, we also note that, if the local GND density $\kappa(\vec r,t)$ is equal to zero, correlations $d^{++}$ and $d^{--}$ display the following symmetry :
\begin{widetext}
\begin{equation}
\label{eq:correlationsymmetry3}
\kappa(\vec r,t) = 0 \quad\rightarrow\quad d^{++}\!(x-x',y-y',\{\rho^s(\vec r,t)\},\tau_{lo}(\vec{r},t))=d^{--}(x-x',y-y',\{\rho^s(\vec r,t)\},\tau_{lo}(\vec{r},t)).
\end{equation}
\end{widetext}

Using the symmetry properties given in Eq.~(\ref{eq:defsym3AF}), it is straightforward to show that the local stresses $\tau^s_f$ and $\tau^s_b$ defined in Eqs.~(\ref{eq:AFdeftauf}) and (\ref{eq:AFdeftaub}) display the following properties:
\begin{widetext}
\begin{eqnarray}
\label{eq:AFprop1}
\tau^s_f(\vec r,\{\rho^s(\vec r,t)\},-\tau_{lo}(\vec r,t))&=&-\tau^s_f(\vec r,\{\rho^s(\vec r,t)\},\tau_{lo}(\vec r,t)) \\
\label{eq:AFprop2}
\tau^s_b(\vec r,\{\rho^s(\vec r,t)\},-\tau_{lo}(\vec r,t))&=&\tau^s_b(\vec r,\{\rho^s(\vec r,t)\},\tau_{lo}(\vec r,t))
\end{eqnarray}
\end{widetext}
where the one-body dislocation densities $\{\rho^s(\vec r,t)\}$ and  local stress $\tau_{lo}(\vec r,t)$ dependencies have been explicitly added and the explicit time dependence suppressed, because $\tau_f^s$ and $\tau_b^s$ inherit this time dependence precisely through $\tau_{lo}(\vec r,t)$ and $\{\rho^s(\vec r,t)\}$. These properties clarify the physical meaning of the local stresses $\tau^s_f$ and $\tau^s_b$. The stresses $\tau^s_f$ change their signs with the sign of the local low-order stress $\tau_{lo}$ and, as shown below in section \ref{numerics}, they are positive when $\tau_{lo}$ is positive. In contrast, the stresses $\tau_b^s$ are invariant upon a change of sign of $\tau_{lo}$. As a consequence, the stresses $\tau_f^s$, which always oppose the low-order stress $\tau_{lo}=\tau_{ext}+\tau_{sc}$ (see Eq.~(\ref{eq:AFneweqq})), play the role of friction stresses whereas the stresses $\tau^s_b$, which are invariant upon a reversal of the local stress $\tau_{lo}$, may generate a Bauschinger effect and a translation of the elastic domain. Therefore, the stresses $\tau^s_b$ play the role of back-stresses.

\section{Broken symmetry in the kinetics of the coarse-grained signed dislocation densities}

It is important to note that, according to Eq.~(\ref{eq:AFneweqq}), the local stress fields experienced respectively by the positive and negative dislocation densities are different: the correlation-induced stress components $\tau^s_f$ and $\tau_b^s$ depend on the sign $s$. In other words, the symmetry that exists at the discrete scale (positive and negative discrete dislocations at the same point $\vec r$ have opposite velocities) is broken at mesoscale: the velocities of positive and negative dislocation densities are not simply of opposite sign. This broken symmetry is the direct consequence of a mesoscale description and its associated coarse-graining procedure: the averaging process required to build a continuous description generates kinetic equations for one-body densities that inevitably incorporate two-body correlations which, in all generality, break the lower-scale symmetry.

In order to be more specific, we analyse explicitly the friction stresses $\tau_f^+$ and $\tau_f^-$ experienced by the positive and negative dislocation densities, respectively. According to Eqs.~(\ref{eq:AFdeftauf}), we have:
\begin{widetext}
\begin{eqnarray}
\label{eq:AFtaufplus31}
\tau^+_f(\vec r)=-\rho^+(\vec r)\int\limits_{\vec r\,'\neq\vec r}\tau_{ind}(\vec r-\vec r\,')\ d^{++}(\vec r-\vec r\,')\ d\vec r\,'+\rho^-(\vec r)\int\limits_{\vec r\,'\neq\vec r}\tau_{ind}(\vec r-\vec r\,')\ d^{+-}(\vec r-\vec r\,')\ d\vec r\,' \\
\label{eq:AFtaufmoins31}
\tau^-_f(\vec r)=-\rho^+(\vec r)\int\limits_{\vec r\,'\neq\vec r}\tau_{ind}(\vec r-\vec r\,')\ d^{-+}(\vec r-\vec r\,')\ d\vec r\,' + \rho^-(\vec r)\int\limits_{\vec r\,'\neq\vec r}\tau_{ind}(\vec r-\vec r\,')\ d^{--}(\vec r-\vec r\,')\ d\vec r\,'
\end{eqnarray}
\end{widetext}
where, because they are not needed for the present argument, the low-order stress and dislocation density dependencies of the correlations and friction stresses have been omitted, as well as the time dependencies.
Using the symmetry property given in Eq.~(\ref{eq:defsym4AF}), it is easy to show that the terms that depend on $d^{++}$ and $d^{--}$ are equal to zero. Therefore, the previous equations reduce to:
\begin{widetext}
\begin{eqnarray}
\label{eq:AFtaufplus312}
\tau^+_f(\vec r)&=&\rho^-(\vec r)\int\limits_{\vec r\,'\neq\vec r}\tau_{ind}(\vec r-\vec r\,')\ d^{+-}(\vec r-\vec r\,')\ d\vec r\,' \\
\label{eq:AFtaufmoins312}
\tau^-_f(\vec r)&=&-\rho^+(\vec r)\int\limits_{\vec r\,'\neq\vec r}\tau_{ind}(\vec r-\vec r\,')\ d^{-+}(\vec r-\vec r\,')\ d\vec r\,'.
\end{eqnarray}
\end{widetext}
Again, using the symmetry properties of Eq.~(\ref{eq:defsym4AF}), it is easy to show that the integrals in Eqs.~(\ref{eq:AFtaufplus312}) and (\ref{eq:AFtaufmoins312}) differ only by their sign. Thus, we have:
\begin{eqnarray}
\label{eq:AFtaufplus313}
\tau^+_f(\vec r)&=&\rho^-(\vec r)A(\vec r) \\
\label{eq:AFtaufmoins313}
\tau^-_f(\vec r)&=&\rho^+(\vec r)A(\vec r)
\end{eqnarray}
with
\begin{equation}
\label{eq:AFdefdeAcorre}
A(\vec r)=\int\limits_{\vec r\,'\neq\vec r}\tau_{ind}(\vec r-\vec r\,')\ d^{+-}(\vec r-\vec r\,')\ d\vec r\,'.
\end{equation}
When the signed densities $\rho^+(\vec r)$ and $\rho^-(\vec r)$ are different, which is the generic situation, the friction stresses $\tau^+_f$ and $\tau^-_f$ are different, which is sufficient to break the symmetry between the velocities of the positive and negative dislocation densities. To better understand this broken symmetry in physical terms, we note that $\rho^-(\vec r)d^{+-}(\vec r\!-\!\vec r\,')$ may be interpreted as the excess (with respect to the uncorrelated state) of negative dislocations in the surrounding of a positive dislocation that sits at point $\vec r$. Equation (\ref{eq:AFtaufplus312}) tells us that this excess of negative dislocations at $\vec r$ is at the origin of the friction stress $\tau^+_f$ experienced by a positive dislocation. There is of course no reason for this excess of negative dislocations around a positive dislocation to be exactly the opposite of the excess of positive dislocations around a negative one. Therefore, the friction stresses $\tau^+_f$ and $\tau^-_f$ ought to be different.\footnote{In fact, this broken symmetry could already have been pointed out earlier when we wrote the coarse-grained kinetic equations in the form of equation (\ref{eq:eqdensiteaveccorrefactorise}), where the dislocation-induced stress was split into the mean field stress $\tau_{sc}(\vec r)$ and the correlation-induced stress $\tau^s_{corr}(\vec r,t)$ defined in Eq.~(\ref{eq:corredetaildependence}). Using the fact that the stress function $\tau_{ind}(\vec r)$ is odd with respect to the coordinate $x$ (see Eq.~(\ref{eq:defAFdetauind})), it is easy to realize that $\tau^s_{corr}(\vec r,t)$ would be independent of $s$ if and only if the correlations $d^{ss'}$ are such that the product $\rho^{s'}(\vec r\,',t)d^{ss'}\!(\vec r,\vec r\,',t)$ is equal to the opposite of $\rho^{\bar s'}(\vec r\,'',t)d^{\bar s\bar s'}\!(\vec r,\vec r\,'',t)$, where $\vec r\,''$ and $\vec r\,'$ are symmetric points with respect to $\vec r$. There is of course absolutely no reason for this to be fulfilled, even if, due to the short range nature of the correlations, point $\vec r\,''$ and $\vec r\,'$ may be restricted to be very close to each other.}

Now, to better visualize this broken symmetry in the signed kinetic equations, we introduce the half sums and half differences of the friction and back-stresses:
\begin{equation}
\label{eq:AFdifecrit}
\begin{matrix}
\tau_f(\vec r)=(\tau_f^+(\vec r)+\tau_f^-(\vec r))/2
\\
\tilde\tau_f(\vec r)=(\tau_f^+(\vec r)-\tau_f^-(\vec r))/2
\\
\tau_b(\vec r)=(\tau_b^+(\vec r)+\tau_b^-(\vec r))/2
\\
\tilde\tau_b(\vec r)=(\tau_b^+(\vec r)-\tau_b^-(\vec r))/2
\end{matrix}
\end{equation}
Using Eqs.~(\ref{eq:AFdeftauf}) and (\ref{eq:AFdeftaub}), we see that these stresses are linked to the correlations $d^{ss'}(\vec r-\vec r\,')$ as follows:
\begin{widetext}
\begin{eqnarray}
\label{eq:AFdefdentf}
\tau_f(\vec r)&=&\frac{1}{2}\rho(\vec r)\int\limits_{\vec r\,'\neq\vec r}\tau_{ind}(\vec r-\vec r\,')\ d^{+-}(\vec r\,-\vec r\,')\ d\vec r\,', \\
\label{eq:AFdefdentb}
\tau_b(\vec r)&=&-\frac{1}{4}\frac{\partial\kappa(\vec r)}{\partial\vec r}\int\limits_{\vec r\,'\neq\vec r}(\vec r\,'-\vec r)\ \tau_{ind}(\vec r-\vec r\,')\left\{d^{++}(\vec r\,'-\vec r)+d^{--}(\vec r\,'-\vec r)+d^{-+}(\vec r\,'-\vec r)+d^{+-}(\vec r\,'-\vec r)\right\}d\vec r\,'  \nonumber \\ 
&&-\frac{1}{4}\frac{\partial\rho(\vec r)}{\partial\vec r}\int\limits_{\vec r\,'\neq\vec r}(\vec r\,'-\vec r)\ \tau_{ind}(\vec r-\vec r\,')\left\{d^{++}(\vec r\,'-\vec r)-d^{--}(\vec r\,'-\vec r)\right\}d\vec r\,' ,\\   
\label{eq:AFdefdenttf}
\tilde\tau_f(\vec r)&=&-\frac{1}{2}\kappa(\vec r)\int\limits_{\vec r\,'\neq\vec r}\tau_{ind}(\vec r-\vec r\,')\ d^{+-}(\vec r-\vec r\,')d\vec r\,',   \\
\label{eq:AFdefdenttb}
\tilde\tau_b(\vec r)&=&-\frac{1}{4}\frac{\partial\kappa(\vec r)}{\partial\vec r}\int\limits_{\vec r\,'\neq\vec r}(\vec r\,'-\vec r)\ \tau_{ind}(\vec r-\vec r\,')\left\{d^{++}(\vec r\,'-\vec r)-d^{--}(\vec r\,'-\vec r)\right\}d\vec r\,'\\
&&-\frac{1}{4}\frac{\partial\rho(\vec r)}{\partial\vec r}\int\limits_{\vec r\,'\neq\vec r}(\vec r\,'-\vec r)\ \tau_{ind}(\vec r-\vec r\,')\left\{d^{++}(\vec r\,'-\vec r)+d^{--}(\vec r\,'-\vec r)-d^{+-}(\vec r\,'-\vec r)-d^{-+}(\vec r\,'-\vec r)\right\}d\vec r\,'   \nonumber
\end{eqnarray}
\end{widetext}
where $\kappa(\vec r)$ is the GND density defined in Eq.~(\ref{eq:AFdefdekapp}) and $\rho(\vec r)$ the total dislocation density:
\begin{equation}
\label{eq:AFdeftotdesn}
\rho(\vec r)=\sum_s\rho^s(\vec r).
\end{equation}
For the sake of compactness of Eqs.~(\ref{eq:AFdefdentf}-\ref{eq:AFdefdenttb}), the dependencies of the correlations on the local dislocation densities and low-order stress $\tau_{lo}(\vec r) = \tau_{ext}+\tau_{sc}(\vec r)$  have been omitted. For latter reference, we note that Eqs.~(\ref{eq:AFdefdentf},\ref{eq:AFdefdenttf}), which implies that $\tilde \tau_f = - \frac{\kappa}{\rho}\tau_f$, together with Eq.~(\ref{eq:AFdifecrit}) lead to the following relation between the sign-dependent friction stresses $\tau_f^s$ and their sign-independent component $\tau_f$:
\begin{equation}
\label{eq:taufs_tauf}
s=\pm 1  \quad : \quad  \tau_f^s(\vec r)=2 \,\frac{\rho^s(\vec r)}{\rho(\vec r)} \, \tau_f(\vec r).
\end{equation}

By definition, $\tau_f(\vec r)$ and $\tau_b(\vec r)$ are the components of the friction and back-stresses experienced by a dislocation independently of its sign, whereas $\tilde\tau_f(\vec r)$ and $\tilde\tau_b(\vec r)$ are their symmetry-breaking counterparts. Indeed, using these stresses, Eqs.~(\ref{eq:AFneweqq}) become
\begin{widetext}
\begin{eqnarray}
\label{eq:AFnewecritdeeqtransplus}
-\frac{\partial\rho^+(\vec r)}{\partial t}&=&+M\vec b  \cdot  \frac{\partial}{\partial\vec r}\left[\rho^+(\vec r)\left\{\tau_{ext}+\tau_{sc}(\vec r)-\tau_f(\vec r)-\tau_b(\vec r)-\tilde\tau_f(\vec r)-\tilde\tau_b(\vec r)\right\}\right] \\
\label{eq:AFnewecritdeeqtransplus2}
-\frac{\partial\rho^-(\vec r)}{\partial t}&=&-M\vec b \cdot \frac{\partial}{\partial\vec r}\left[\rho^-(\vec r)\left\{\tau_{ext}+\tau_{sc}(\vec r)-\tau_f(\vec r)-\tau_b(\vec r)+\tilde\tau_f(\vec r)+\tilde\tau_b(\vec r)\right\}\right].
\end{eqnarray}
\end{widetext}
where we clearly see that $\tau_f$ and $\tau_b$ drive dislocations with opposite Burgers vector along opposite directions, whereas the symmetry-breaking stresses $\tilde \tau_f$ and $\tilde \tau_b$ drive dislocations of opposite signs along the \emph{same} direction.

Similar equations have already been proposed \cite{ZA2001,2GR2002} (see also footnote \footnote{Recently, M. Geers and coll. \cite{DO2015} have proposed a set of different transport equations based on dislocation densities. However, the derivation does not rely on a coarse-graining procedure and, thus, does not correspond to a real transition to mesoscale.} and Ref. \onlinecite{DO2015}), but without the symmetry-breaking stresses $\tilde\tau_f(\vec r)$ and $\tilde\tau_b(\vec r)$ and with a sign-independent back-stress $\tau_b(\vec r)$ limited to the term that depends on the gradient of the polar (GND) density $\kappa(\vec r)$, i.e. to the $1^{\text{st}}$ term in the right hand side of Eq.~(\ref{eq:AFdefdentb}). 

Finally, we mention that, in the limit $\kappa(\vec r)\! \ll \! \rho(\vec r)$, the back-stresses that enter into kinetic equations (\ref{eq:AFnewecritdeeqtransplus}) and (\ref{eq:AFnewecritdeeqtransplus2}) may be simplified. More precisely, using the fact the difference $d^{++}(\vec r\,'-\vec r)-d^{--}(\vec r\,'-\vec r)$ is, to the lowest order, linear in $\kappa(\vec r)/\rho(\vec r)$ (consequence of the property given in Eq.~(\ref{eq:correlationsymmetry3})), an analysis of the kinetic equations, to the lowest order in fluctuations of the dislocation densities around an homogeneous state with no GND, shows that we can neglect the terms that depend on the difference $(d^{++}-d^{--})$ and approximate the back-stresses by:

\begin{widetext}
\begin{eqnarray}
\label{eq:AFdefdentbmodreduced}
\tau_b(\vec r)&\simeq&-\frac{1}{4}\frac{\partial\kappa}{\partial\vec r}\int\limits_{\vec r\,'\neq\vec r}(\vec r\,'-\vec r)\ \tau_{ind}(\vec r-\vec r\,')\left\{d^{++}(\vec r\,'-\vec r)+d^{--}(\vec r\,'-\vec r)+d^{-+}(\vec r\,'-\vec r)+d^{+-}(\vec r\,'-\vec r)\right\}d\vec r\,'  ,\\   
\label{eq:AFdefdenttbreduced}
\tilde\tau_b(\vec r)&\simeq&-\frac{1}{4}\frac{\partial\rho}{\partial\vec r}\int\limits_{\vec r\,'\neq\vec r}(\vec r\,'-\vec r)\ \tau_{ind}(\vec r-\vec r\,')\left\{d^{++}(\vec r\,'-\vec r)+d^{--}(\vec r\,'-\vec r)-d^{+-}(\vec r\,'-\vec r)-d^{-+}(\vec r\,'-\vec r)\right\}d\vec r\,'   
\end{eqnarray}
\end{widetext}
Of course, these approximations are valid provided the kinetics preserve the constraint $\kappa(\vec r)\! \ll \! \rho(\vec r)$, which is certainly not a generic situation, in particular in situations where the plastic strain develops strong heterogeneities.

\section{Numerical coarse-graining procedure\label{numerics}}

Transport equations of a mesoscale dislocation density theory contain correlation induced stresses, specifically friction and back-stress terms.
These terms depend on the correlation functions $d^{ss'\!}$, which must be computed through a coarse-graining procedure. As explained in section~\ref{correldeAF}, if the width $L$ of the spatial convolution window is large enough and the time convolution window appropriately chosen, the correlations $d^{ss'}\!(\vec r,\vec r\,',t)$ may be considered as functions of $(\vec r\!-\!\vec r\,')$ and of the local densities $\rho^s(\vec r,t)$. 

We focus here on the sign-independent friction stress $\tau_f$ defined in Eq.~(\ref{eq:AFdefdentf}). Due to the local character of the correlations, which is a direct consequence of the underlying coarse-graining procedure, Eq.~(\ref{eq:AFdefdentf}) may be written as:
\begin{eqnarray}
\label{eq:taufpourrhoprhomadimpourcorrechapde}
\tau_f(\vec{r})&=&\frac{1}{2}\, \rho(\vec{r}) \int\limits_{(x,y)\neq(0,0)}  \tau_{ind}(x,y) \nonumber  \\
&\times& d^{+-}\left({x},{y},\rho(\vec r),\kappa(\vec r),\tau_{lo}(\vec r),L\right)\ d{x}d{y}
\end{eqnarray}
where the origin of the coordinates $(x,y)$ is located at point $\vec r$. The dependencies of the correlations with the local one-boby densities and low-order stress $\tau_{lo}(\vec r) = \tau_{ext} + \tau_{sc}(\vec r)$, sum of the applied stress and  long-ranged mean-field stress, have been reintroduced.  A $L$ dependency has been also explicitly pointed out because the length $L$, together with the associated time window $T(L)$ and the statistical ensemble of initial conditions, characterises the coarse-graining procedure used to define the mesoscale one- and two-body dislocation densities and, consequently, the correlations.

Here, we recourse to 2D Discrete Dislocation Dynamics (DDD) to compute numerically the correlations. In principle, for a given coarse-graining length $L$, correlations at point $\vec r$ and their variations with the local dislocation densities $\kappa(\vec r)$ and $\rho(\vec r)$ and the low-order stress $\tau_{lo}(\vec r)$ should be analysed in the context of a system whose linear dimensions are much larger than $L$, keeping in mind that the dislocation densities should still be defined and homogeneous at scale $L$. Due to their local character and short-range nature, correlations $d^{ss'}$ in the neighborhood of point $\vec r$ depend only on the local values of the one-body densities $\kappa(\vec r)$ and $\rho(\vec r)$ (see Eq.~(\ref{eq:defdeffefAF2})). We may therefore consider a situation where the densities $\kappa(\vec r)$ and $\rho(\vec r)$ are uniform within the system and equal to the values we want to investigate. In that case, due to the symmetry property of $\tau_{ind}(\vec r)$, see Eq.~(\ref{eq:defAFdetauind}), the self-consistent stress $\tau_{sc}(\vec r)$ vanishes and the low-order stress $\tau_{lo}(\vec r)$ is simply equal to the applied stress $\tau_{ext}$. Next, using again the fact that the correlation length is of the order of the average dislocation spacing $1/\sqrt{\rho}$, we may safely replace the large system by a minimal finite box of linear dimension equal to the coarse-graining length $L$ , supplemented by periodic boundary conditions, provided of course $L$ is sufficiently larger the $1/\sqrt{\rho}$.

As a result, the spatial convolution window is simply a constant window function of size $L$, the linear size of the DDD simulation box. $L$ should be of the order of the spatial resolution of the continuous model we want to develop and, as just recalled, sufficiently larger than $1/\sqrt\rho$, the average distance between dislocations. This guaranties that $L$ will always be significantly larger than the range of the correlations $d^{ss'}$. As explained in section~\ref{correldeAF}, the relevant choice for the time window, that in all generality should depend on $L$, is to select $T(L)$ of the order of the average time needed by the dislocations to reach a stationary or a steady state, depending on whether the dislocations adopt a quasi-static or a flowing state. This guaranties that $T(L)$ is long enough but still smaller than the characteristic time of the evolution of the one-body densities. Finally, this space and time convolution is supplemented by a statistical average over an ensemble of random initial dislocation configurations, as explained in section~\ref{subsectionCG}. In line with the argument developed there, which states that no more information than the one embedded in the initial one-body dislocation densities should be used, this statistical ensemble should simply be defined by uniform distribution functions $f_s$. As the mesoscopic densities read $\rho^s=N^s/L^2$, where $N^s$ is the number of dislocations of sign $s$, and taking into account that the spatial convolution window is constant within the simulation box, Eq.~(\ref{eq:AFbecome2})  leads simply to $f_{+}=f_{-} = 1/L^2$.

Prior to its numerical analysis, we exhibit the scaling behaviour of the friction stress. We note that the dislocation kinetics given in Eq.~(\ref{eq:vitessedislodetail}) is invariant upon rescaling the lengths by $1/\sqrt{\rho}$, the applied stress by $\mu b \sqrt{\rho}/2\pi(1-\nu)$ and the time by $2\pi(1-\nu)/\rho M \mu b^2$, where $\rho$ is the total dislocation density. We naturally extend this rescaling to the choice of the spatial and temporal widths $L$ and $T(L)$  of the coarse-graining convolution window $w(\vec r,t)$ defined in Eq.~(\ref{eq:AFfuncw}). Hence, the overall scale invariance of the kinetics and of the coarse-graining procedure implies that the correlations follow scaling forms
\begin{eqnarray}
\label{eq:scalingform}
 d^{ss'}&&\left({x},{y},\rho,\kappa,\tau,L\right)= \nonumber \\
 &&f^{s,s'}\left(x\sqrt{\rho},y\sqrt{\rho},\frac{\kappa}{\rho},\frac{2\pi(1-\nu)\tau_{ext}}{\mu b\sqrt{\rho}},L\sqrt\rho\right)\end{eqnarray}
where, as we consider here a single finite system of linear size $L$, there is no need to specify a $\vec r$ dependence of the local mesoscopic quantities. This scale invariance, in turn, implies that the friction stress given in Eq.~(\ref{eq:taufpourrhoprhomadimpourcorrechapde}) follows the scaling form 
\begin{equation}
\label{eq:scalingformtauf1}
\tau_f =\ \frac{\mu b \, \sqrt{\rho} }{2\pi(1-\nu)}\; f\left(\frac{\kappa}{\rho},\frac{2\pi(1-\nu)\tau_{ext}}{\mu b\sqrt{\rho}},L\sqrt\rho\right)
\end{equation}
where the scaling function $f$ is defined by :
\begin{eqnarray}
\label{eq:scalingformtauf2}
f&&\left(\frac{\kappa}{\rho},\frac{2\pi(1-\nu)\tau_{ext}}{\mu b\sqrt{\rho}},L\sqrt\rho\right) =\frac{1}{2}\ \int\limits_{(\tilde{x},\tilde{y})\ne(0,0)}\frac{\tilde{x}(\tilde{x}^2-\tilde{y}^2)}{{(\tilde{x}^2+\tilde{y}^2)}^2} \nonumber \\
&&\times f^{+-}\left(\tilde{x},\tilde{y},\frac{\kappa}{\rho},\frac{2\pi(1-\nu)\tau_{ext}}{\mu b\sqrt{\rho}},L\sqrt\rho\right) \ d\tilde{x}d\tilde{y}
\end{eqnarray}
where $\tilde{x}$ and $\tilde{y}$ are the cartesian coordinates in units of $1/\sqrt{\rho}$. We note that, because of the $\rho$-dependence of the scaling function $f$, the friction stress does not simply scaled as $\sqrt{\rho}$.

The coarse-grained scaling function $f$ needs now to be estimated numerically. 

Generally speaking, we may expect that the coarse-graining length $L$ will show up in the coarse-grained quantities that result from the coarse-graining procedure. The important point is that we are dealing here with a situation where many length scales may emerge from the complex spatial and dynamical coupling that governs the dislocation dynamics. It is indeed well known that, most often, dislocations self-organized themselves into complex patterns that display length scales much larger than the average dislocation spacing, such as dislocation walls in cyclic loading \cite{LE2004} or even seemingly fractal structures \cite{MU1986}  with no characteristic length scale \cite{HA1998,ZA1999}. In such situations, when many different large length scales are physically present, an averaging procedure at a given intermediate length scale will generate a continuous theory which is scale dependent. In the present context, it means that the correlation-induced stresses generated by coarse-graining may definitely display an $L$-dependence.

\begin{figure}[h]
\centering
\includegraphics[width=1.0\linewidth]{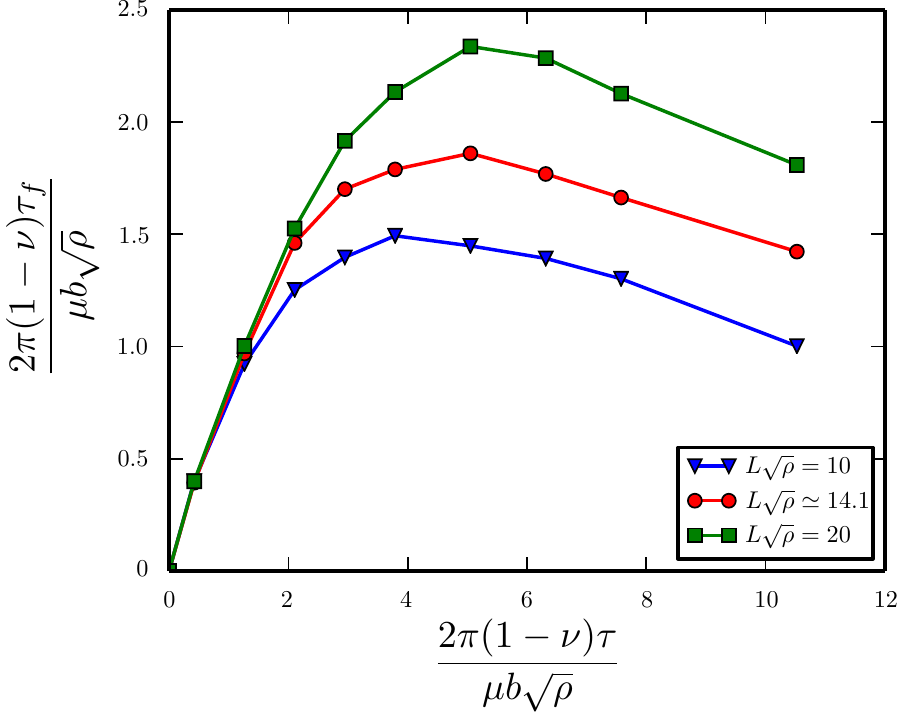}
\caption{Numerical results for the friction stress $\tau_f$ as a function of the applied stress and for different dimensionless coarse-graining length $L\sqrt{\rho}$.}
\label{fig:numericrestf}
\end{figure}

Therefore, in order to investigate this important feature, we consider below different values of $L$. In fact, as the only pertinent quantity is $L\sqrt\rho$, we analyse different values of $\sqrt N\!=\!L\sqrt\rho$, where $N$ is the total number of dislocations. The analysis is restricted to situations where the number of positive and negative dislocations are equal. Therefore, the GND density $\kappa$ is set to zero and the computations are performed for different values of the applied stress. The results for three different values of the parameter $\sqrt N\!=\!L\sqrt\rho$ are presented in Fig.~\ref{fig:numericrestf}.

First, we observe that the stress $\tau_f$ is positive when the applied stress $\tau_{ext}$ is positive. This property could have been qualitatively anticipated. Indeed, when the applied stress in non zero, the average $45^{\circ}$-alignment of the short-ranged dipoles, observed in the absence of applied stress, is modified: a simple analysis of the profile along the glide direction $x$ of the dislocation-dislocation interaction $\tau_{ind}(x,y)$ given in Eq.~(\ref{eq:defAFdetauind}) shows that, for $\tau_{ext}>0$, the correlation function $d^{+-}(x,y)$ (which is proportional to the excess probability of having a positive dislocation at $(x,y)$ if a negative one sits at the origin) displays maxima $(x_m,y_m)$ characterised by $\lvert x_m \rvert \!<\! \lvert y_m \rvert$ (respectively, $x_m \!>\! \lvert y_m \rvert$) in the half-plane $x\!<\!0$ (respectively, $x\!>\!0$). These maxima lie in regions where the function $\tau_{ind}(x,y)$ is positive. This makes the integral that enters the r.h.s of Eq.~(\ref{eq:taufpourrhoprhomadimpourcorrechapde}) positive. Therefore, the correlation-induced stress $\tau_f$ should be positive when the applied stress is positive. This is indeed what we observe in Fig.~\ref{fig:numericrestf}.  Now, we note that, according to Eq.~(\ref{eq:taufs_tauf}), the sign-dependent stresses $\tau_f^s$ ($s=\pm 1$) and $\tau_f$ have the same sign. In conclusion, as stated in section \ref{taufs_tau_bs}, the stresses $\tau_f^s$ are positive when the local low-order stress $\tau_{lo}$ (here reduced to $\tau_{ext}$) is positive and they change their signs with the sign of $\tau_{lo}$. In other words, the stresses $\tau_f^s$ always oppose $\tau_{lo}$ : they act as friction terms. 

Second, we observe that the friction stress $\tau_f$ vanishes with the applied stress $\tau_{ext}$ and decreases for large $\tau_{ext}$. These limits are in fact easily predictable. First, when $\tau_{ext}$ is equal to zero, correlations $d^{+-}$, and therefore their scaling form $f^{+-}$, display an axial symmetry with respect to the $y$-axis. Consequently, $f$, which is the integral of an odd function (see Eq.~(\ref{eq:scalingformtauf2})) is equal to zero, which implies that $\tau_f$ is also equal to zero. Second, when the stress $\tau_{ext}$ is large enough, the individual dislocation-dislocation interactions become negligible compared to $\tau_{ext}$. Consequently, dislocations with opposite Burgers vectors become less correlated contrary to dislocations of the same sign. Therefore, when $\tau_{ext}$ is large enough, the amplitude of the correlations $d^{+-}$  decreases when $\tau_{ext}$ increases and, consequently, the friction stress $\tau_f$ also decreases.

In fact, the friction stress displays two different regimes. For small applied stresses (up to approximately $0.3$ in dimensionless units), the friction term is approximatively linear with a slope close to $1$. Therefore, the friction term opposes almost totally the applied stress. This is associated to a quasi-static state where there is no effective dislocation flow. For higher applied stresses, the friction stress becomes smaller than the applied stress. This regime is associated to a permanent dislocation flow. This behaviour is in agreement with the direct observation of the DDD simulations.

\begin{figure}[h]
\centering
\includegraphics[width=1.0\linewidth]{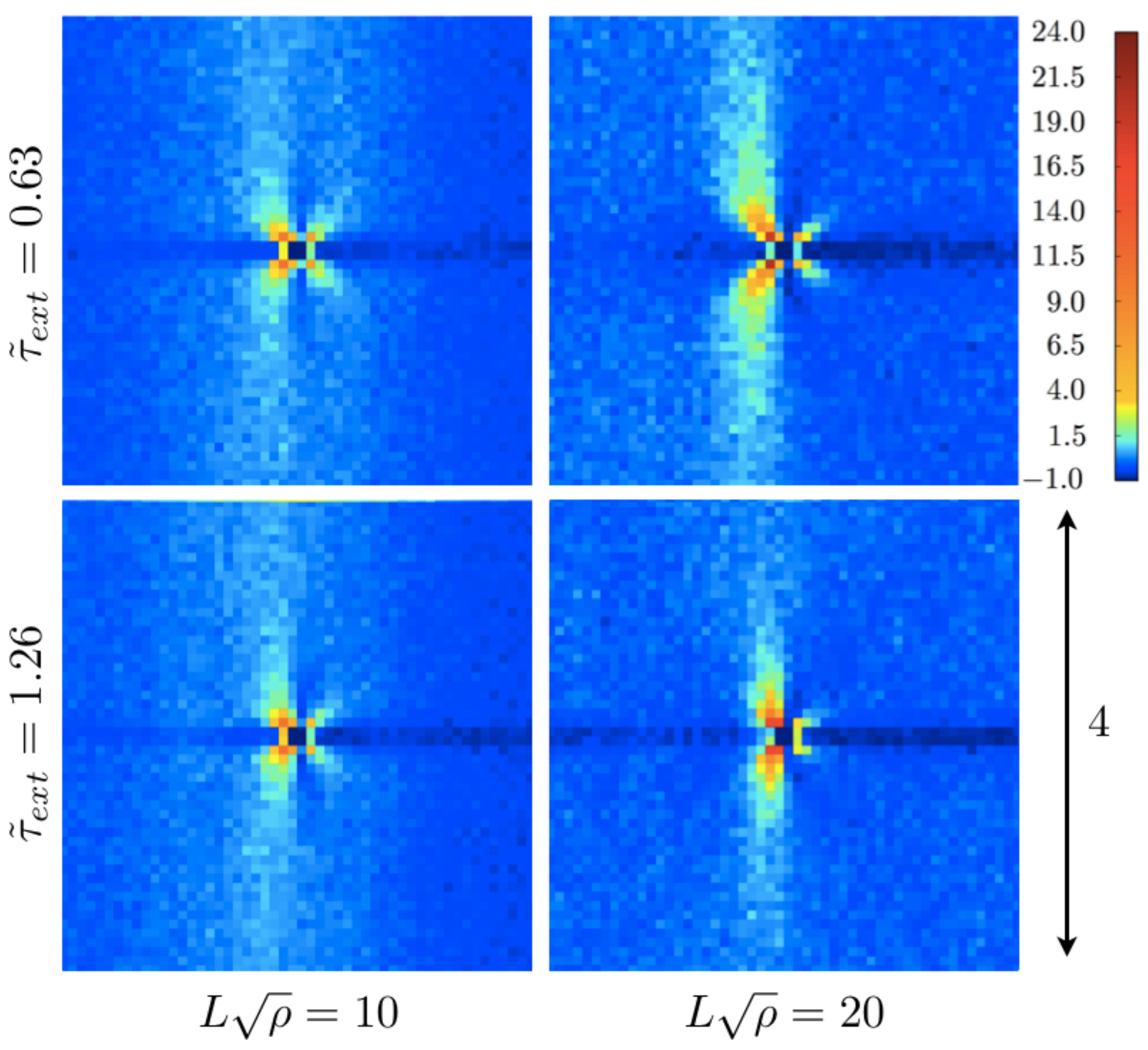}
\caption{Zooms of correlation maps $d^{+-}$ for different sizes of the total simulation box. Left column: $L_x\sqrt{\rho}\!=\!10$ and $L_y\sqrt{\rho}\!=\!20$,  right column: $ L_x\sqrt{\rho}\!=\!L_y\sqrt{\rho}\!=\!20$. Two different values of the applied stress $\tilde \tau_{ext} \!=\! 2\pi (1-\nu) \tau_{ext} /( \mu b \sqrt{\rho})$ are considered : top row, $\tilde \tau_{ext}\!=\!0.63$; bottom row, $\tilde \tau_{ext}\!=\!1.26$. The linear dimension of the zooms is $d\sqrt{\rho}\!=4$.}
\label{fig:influxsurlescorre}
\end{figure}

Now, we comment on the dependence of the friction stress $\tau_f$ with the parameter $L\sqrt\rho$. Figure~\ref{fig:numericrestf} shows that, for a given density $\rho$, the stress is scale dependent. In light of the previous discussion, this is not surprising. Examination of the simulated dislocation configurations indicates that this is due to the increase with $L$ of the number of very short-range dipoles formed by two dislocations of opposite sign. This is quantitatively confirmed by the correlation maps (see Fig.~\ref{fig:influxsurlescorre}), where we observe that the correlation function $d^{+-}$, in a very close neighborhood of the origin, increases significantly when we double the size along $x$ of the simulation box, keeping the same density $\rho$.

The physical origin of the increase of the number of dipoles with $L$ (at constant dislocation density) is that the coarse-graining procedure involves a time convolution with a temporal width $T(L)$ that, when the dislocations adopt a flowing state, is of the order of the travelling time over the length $L$. Therefore, the probability that a given dislocation meets another dislocation of opposite sign during the time $T(L)$ increases with $L$. In brief, the longer $L$, the higher the number of dipole that have the time to form. However, we note that this physical phenomena may be here disturbed by the use of periodic boundary conditions because a dislocation may travel through the simulation box more than once. This undesirable effect may be avoided with a careful numerical monitoring of $T(L)$, which has not been done here. Therefore, the $L$-dependence observed in Fig.~\ref{fig:numericrestf}, even if it has a true physical origin, may not be perfectly quantitative.

Before to conclude, we briefly extend to all the correlation-induced stresses the scaling form presented above for the sign-independent friction stress $\tau_f$. 

Using the scaling forms of the correlations $d^{ss'}$ given in Eq.~(\ref{eq:scalingform}), the correlation-induced stresses given in Eqs.~(\ref{eq:AFdefdentf}-\ref{eq:AFdefdenttb}) adopt the following scaling forms 
\begin{widetext}
\begin{eqnarray}
\label{eq:scaling_tau_f}
\tau_f &=& Gb\;\sqrt{\rho} \; \; f\left(\frac{\kappa}{\rho},\frac{\tau_{ext}}{G b\sqrt{\rho}},L\sqrt\rho\right) \\
\label{eq:scaling_taub}
\tau_b &=& Gb\;\; \frac{\kappa}{\rho^2}\;\frac{\partial \,\rho}{\partial x} \;\;h\left(\frac{\tau_{ext}}{G b\sqrt{\rho}},L\sqrt\rho\right) + Gb \;\frac{1}{\rho}\;\frac{\partial \,\kappa}{\partial x} \;\left\{C_{++}\left(\frac{\kappa}{\rho},\frac{\tau_{ext}}{G b\sqrt{\rho}},L\sqrt\rho\right)-C_{+-}\left(\frac{\kappa}{\rho},\frac{\tau_{ext}}{G b\sqrt{\rho}},L\sqrt\rho\right)\right\}\\
\label{eq:scaling_tilde_tau_f}
\tilde \tau_f &=& -Gb\;\;\frac{\kappa}{\rho}\;\sqrt{\rho} \; \; f\left(\frac{\kappa}{\rho},\frac{\tau_{ext}}{G b\sqrt{\rho}},L\sqrt\rho\right) \\
\label{eq:scaling_tilde_tau_b}
\tilde \tau_b &=& Gb\;\;\frac{\kappa}{\rho^2}\;\frac{\partial\, \kappa}{\partial x} \; \;h\left(\frac{\tau_{ext}}{G b\sqrt{\rho}},L\sqrt\rho\right) + Gb  \;\frac{1}{\rho}\;\frac{\partial \,\rho}{\partial x} \;\left\{C_{++}\left(\frac{\kappa}{\rho},\frac{\tau_{ext}}{G b\sqrt{\rho}},L\sqrt\rho\right)+C_{+-}\left(\frac{\kappa}{\rho},\frac{\tau_{ext}}{G b\sqrt{\rho}},L\sqrt\rho\right)\right\}
\end{eqnarray}
\end{widetext}
where, for simplicity, the $\vec r$ dependencies of the local stresses and dislocation densities have been omitted. Function $f$ has been given above in Eq.~(\ref{eq:scalingformtauf2}). The scaling functions $C_{++}$ and $C_{+-}$ are given by
\begin{widetext}
\begin{eqnarray}
\label{eq:defdecplusplus}
C_{++}\left(\frac{\kappa}{\rho},\frac{\tau_{ext}}{G b\sqrt{\rho}},L\sqrt\rho\right)&=&+\frac{1}{2}\ \int\limits_{(\tilde{x},\tilde{y})\ne(0,0)}\frac{\tilde{x}^2(\tilde{x}^2-\tilde{y}^2)}{{(\tilde{x}^2+\tilde{y}^2)}^2} \; f^{++}\left(\tilde{x},\tilde{y},\frac{\kappa}{\rho},\frac{\tau_{ext}}{G b\sqrt{\rho}},L\sqrt\rho\right) \ d\tilde{x}d\tilde{y} \\
\label{eq:defdecplusmoins}
C_{+-}\left(\frac{\kappa}{\rho},\frac{\tau_{ext}}{G b\sqrt{\rho}},L\sqrt\rho\right)&=&-\frac{1}{2}\ \int\limits_{(\tilde{x},\tilde{y})\ne(0,0)}\frac{\tilde{x}^2(\tilde{x}^2-\tilde{y}^2)}{{(\tilde{x}^2+\tilde{y}^2)}^2} \; f^{+-}\left(\tilde{x},\tilde{y},\frac{\kappa}{\rho},\frac{\tau_{ext}}{G b\sqrt{\rho}},L\sqrt\rho\right) \ d\tilde{x}d\tilde{y}.
\end{eqnarray}
\end{widetext}
The scaling function $h$, which does not depend on the ratio $\kappa/\rho$,  is given by the relation
\begin{equation}
\label{h_hh}
H\left(\frac{\kappa}{\rho},\frac{\tau_{ext}}{G b\sqrt{\rho}},L\sqrt\rho\right) \simeq \frac{\kappa}{\rho}\; h\left(\frac{\tau_{ext}}{G b\sqrt{\rho}},L\sqrt\rho\right)
\end{equation}
where the scaling function $H$ is given by
\begin{widetext}
\begin{eqnarray}
\label{eq:deffunctionh}
H\left(\frac{\kappa}{\rho},\frac{\tau_{ext}}{G b\sqrt{\rho}},L\sqrt\rho\right)=\frac{1}{4}\ \int\limits_{(\tilde{x},\tilde{y})\ne(0,0)}&&\frac{\tilde{x}^2(\tilde{x}^2-\tilde{y}^2)}{{(\tilde{x}^2+\tilde{y}^2)}^2} \; \{f^{++}\left(\tilde{x},\tilde{y},\frac{\kappa}{\rho},\frac{\tau_{ext}}{G b\sqrt{\rho}},L\sqrt\rho\right) \\
&&-\; f^{--}\left(\tilde{x},\tilde{y},\frac{\kappa}{\rho},\frac{\tau_{ext}}{G b\sqrt{\rho}},L\sqrt\rho\right) \} \ d\tilde{x}d\tilde{y}
\end{eqnarray}
\end{widetext}
This approximation used in Eq.~(\ref{h_hh}) results from a first oder expansion in $\kappa/\rho$ of $H$, which, according to the property given in Eq.~(\ref{eq:correlationsymmetry3}), vanishes when  the GND density $\kappa$ vanishes. Functions $f^{ss'}$ that appear in the previous equations are the scaling forms of the correlations $d^{ss'}$, as defined in Eq.~(\ref{eq:scalingform}). The first term on the r.h.s. of Eq.~(\ref{eq:scaling_taub}), which concerns the sign-independent back-stress $\tau_b$, has been recently discussed by T. Hochrainer within the context of a thermodynamics approach of the continuum dislocation dynamics \cite{HO2016} that, in its present form, does not include any reference to the symmetry-breaking stresses $\tilde \tau_f$ and $\tilde \tau_b$ introduced here and given in Eqs.~(\ref{eq:scaling_tilde_tau_f}) and (\ref{eq:scaling_tilde_tau_b}), respectively.

\section{Summary}

We have clarified the mathematical procedure needed to coarse-grain dislocation dynamics from the discrete to the continuum. In particular, we have emphasised that the coarse-graining procedure requires a space and time convolution, supplemented by an average on a statistical ensemble. We also argued that, if the width $L$ of the spatial correlation and the width $T(L)$ of the associated time convolution are both large enough, the mesoscopic two-body correlations may be considered locally invariant by translation and stationary at the scale of the characteristic evolution time of the one-body densities. In other words, we may write $d^{ss'}\!(\vec r,\vec r\,',t)\!\simeq\!d^{ss'}\!(\vec r\!-\!\vec r\,',\{\rho^s(\vec r,t)\})$.

We have explained that the coarse-graining procedure generates correlation-induced stresses $\tau^s_f$ and $\tau_b^s$ that have specific physical interpretations. The stresses $\tau_f^s$ change their signs with the sign of the local low-order stress $\tau_{lo}$ (sum of the applied stress and the mean-field stress) and are positive when $\tau_{lo}$ is positive. Therefore, the stresses $\tau_f^s$ always oppose the local stress $\tau_{lo}$: they act as friction stresses. In contrast, the stresses $\tau^s_b$ are invariant upon a reversal of the local stress $\tau_{lo}$. Therefore, they may generate a Bauschinger effect and a translation of the elastic domain: they act as back-stresses.

The friction and back-stresses $\tau^s_f$ and $\tau_b^s$, that depend on the sign $s$ of the Burgers vector, can be further separated into sign-independent and symmetry-breaking contributions. We have shown that the sign-independent back-stress $\tau_b$, which has been usually limited to a term that depends on the gradient of the GND density, contains also a term that depends on the gradient of the total density.

We have also shown that the symmetry-breaking components of the friction and back-stresses, $\tilde \tau_f$ and $\tilde \tau_b$, break the symmetry of the kinetic equations: they drive dislocations of opposite Burgers vectors along the same direction. In other words, within the mesoscopic transport equations, positive and negative dislocation densities do not experience the same local stress: they display velocities which are not strictly opposite.

Finally, using 2D simulations of the discrete dislocation dynamics, we observed a $L$-dependence of the coarse-grained friction stress. This length-scale dependence is not surprising, regarding the frequently observed patterns that dislocation dynamics often generate. These patterns generally exhibit characteristic length scales much larger than the average distance between dislocations. Therefore, a coarse-graining procedure based on a length scale $L$ smaller than these configurational length scales will inevitable lead to correlation-induced stresses that are $L$-dependent. In the present oversimplified situation, where parallel dislocations are limited to a single glide system, the $L$-dependence has been linked to the dynamical formation of short-range dipoles associated to the spatial and time scales of the coarse-graining procedure.

\small
\bibliographystyle{apsrev4-1.bst}

\begin{thebibliography}{28}%


\bibitem{NY1953} J.F.~Nye, Some geometrical relations in dislocated crystals, Acta Mater. 1, 153 (1953).
\bibitem{KR1959} E.~Kr\"oner, Allgemeine Kontinuumstheorie der Versetzungen und Eigenspannungen, Archive for Rational Mechanics and Analysis 4, 273 (1959).
\bibitem{AC2001} A.~Acharya, A model of crystal plasticity based on the theory of continuously distributed dislocations, J. Mech. Phys. Solids 49, 761 (2001).
\bibitem{AC2003} A.~Acharya, Driving forces and boundary conditions in continuum dislocation mechanics, Proc. R. Soc. A 459, 1343 (2003).
\bibitem{FR2011} C.~Fressengeas, V.~Taupin and L.~Capolungo, An elasto-plastic theory of dislocation and disclination fields, V.~Taupin and L.~Capolungo, Int. J. Solids Struct. 48, 3499 (2011).
\bibitem{FR2014} C.~Fressengeas and V.~Taupin, A field theory of distortion incompatibility for coupled fracture and plasticity, J. Mech. Phys. Solids 68, 45 (2014).
\bibitem{RO2005} A.~Roy and A.~Acharya, A field theory of distortion incompatibility for coupled fracture and plasticity, J. Mech. Phys. Solids 53, 143 (2005).
\bibitem{TAU2013} V.~Taupin, L.~Capolungo, C.~Fressengeas, A.~Das and M.~Upadhyay, A field theory of distortion incompatibility for coupled fracture and plasticity, J. Mech. Phys. Solids 61, 370 (2013).
\bibitem{KO1962} A.M.~Kosevich. Zh. Eksper. Fiz. 42 (1962) 152 (in Russian);  Soviet Phys. JEPT (English transl.) 15 (1962) 108.
\bibitem{MU1963} T.~Mura, On dynamic problems of continuous distribution of dislocations, Int. J. Eng. Sci. 1, 371 (1963).
\bibitem{LA1970} L.D.~Landau and E.M.~Lifshitz, Course of Theoretical Physics: Theory of Elasticity, Pergamon Press (1970).
\bibitem{NA1979} A.M.~Kosevich, Crystal dislocations and the theory of elasticity, in Dislocations in Solids, Ed. F.R.N. Nabarro, North-Holland, Amsterdam, 33 (1979).
\bibitem{AC2006} A.~Acharya and A.~Roy, Size effects and idealized dislocation microstructure at small scales: Predictions of a Phenomenological model of Mesoscopic Field Dislocation Mechanics: Part I, J. Mech. Phys. Solids 54, 1687 (2006).
\bibitem{FR2010} C.~Fressengeas, A.~Acharya and A.J.~Baudoin, Dislocation Mediated Continuum Plasticity: Case Studies on Modeling Scale Dependence, Scale-Invariance, and Directionality of Sharp Yield-Point, Computational Methods for Microstructure-Property Relationship, 277 (2010).
\bibitem{RO2006} A.~Roy and A.~Acharya, Dislocation Mediated Continuum Plasticity: Case Studies on Modeling Scale Dependence, Scale-Invariance, and Directionality of Sharp Yield-Point, J. Mech. Phys. Solids 54, 1711 (2006).
\bibitem{HO2006} T.~Hochrainer, \textit{Evolving Systems of Curved Dislocations: Mathematical Foundations of a Statistical Theory}, (Ph.D. Thesis, 2006).
\bibitem{HO2007} T.~Hochrainer, M.~Zaiser and P.~Gumbsch, A three-dimensional continuum theory of dislocation systems: kinematics and mean-field formulation, Philos. Mag. A 87, 1261 (2007).
\bibitem{SA2011} S.~Sandfeld, T.~Hochrainer, M.~Zaiser and P.~Gumbsch, Continuum modeling of dislocation plasticity: Theory, numerical implementation, and validation by discrete dislocation simulations, J. Mater. Res. 26, 623 (2011).
\bibitem{AZ2000} A.~El-Azab, Statistical mechanics treatment of the evolution of dislocation distributions in single crystals, Phys Rev. B 61, 11956 (2000).
\bibitem{AZ2006} A.~El-Azab, Statistical mechanics of dislocation systems, Scr. Mater. 54, 723 (2006).
\bibitem{1GR1997} I.~Groma, Link between the microscopic and mesoscopic length-scale description of the collective behavior of dislocations, Phys Rev. B 56, 5807 (1997).
\bibitem{ZA2001} M.~Zaiser, M.-C.~Miguel and I.~Groma, Statistical dynamics of dislocation systems: The influence of dislocation-dislocation correlations, Phys Rev. B 64, 224102 (2001).
\bibitem{2GR2002} I.~Groma, F.F.~Csikor and M.~Zaiser, Spatial correlations and higher-order gradient terms in a continuum description of dislocation dynamics, Acta Mater. 51, 1271 (2003).
\bibitem{FI2014} A.~Finel, P.-L.~Valdenaire, Y.~Le~Bouar, B.~Appolaire, Dislocation density model : coarse-graining and correlations, Sch\"{o}ntal Symposium on Dislocation-based Plasticity, 24-28 february 2014, Bad Sch\"{o}ntal, Germany.
\bibitem{MA2003} G.~Mazenko, Fluctuations, Order, and Defects, John Wiley and Sons Ltd (2003).
\bibitem{ZA2015}M.~Zaiser, Local density approximation for the energy functional of three-dimensional dislocation systems, Phys. Rev. B. 92, 174120 (2015)
\bibitem{IS2011} P.D~Isp\'{a}novity, I.~Groma, G.~Gy\"{o}rgyi, P.~Szab\'o, W.~Hoffelner, Criticality of Relaxation in Dislocation Systems,Phys Rev. Lett. 107, 085506 (2011).
\bibitem{IS2014} P. D.~Isp\'{a}novity, L.~Laurson, M.~Zaiser, I.~Groma, S.~Zapperi, M.J.~Alava, Avalanches in 2D dislocation systems: Plastic yielding is not depinning, Phys Rev. Lett. 112, 235501 (2014).
\bibitem{DO2015} M.M.W. ~Dogge, R.H.J.~Peerlings, M.G.D.~Geers, Extended modelling of dislocation transport-fomulation and finite element implementation, Adv. Model. and Simul. in Eng. Sci., 2, 29 (2015) 
\bibitem{LE2004} M.~Legros, A.~Jacques, A.~Georges, Cyclic deformation of silicon single crystals: mechanical behaviour and dislocation arrangements, Mat. Sci. Eng. A 387-389 (2004) 495.
\bibitem{MU1986} H.~Mughrabi, T.~Ungar, W.~Kienle, M.~Wilkens, Long-range internal stresses and asymmetric X-ray line-broadening in tensile-deformed [001]-orientated copper single crystals, Philos. Mag. A 53, 793 (1986).
\bibitem{HA1998}P.~Hahner, K.~Bay, M.~Zaiser, Fractal Dislocation Patterning During Plastic Deformation, Phys. Rev. Lett. 81, 2470 (1998)
\bibitem{ZA1999}M.~Zaiser, K.~Bay, P.~Hahner, Fractal analysis of deformation-induced dislocation patterns, Acta Metallurgica, 47, 2463 (1999).
\bibitem{HO2016} T.~Hochrainer, Thermodynamically consistent continuum dislocation dynamics, J. Mech. Phys. Solids 88, 12 (2016).









\end{thebibliography}


%

\end{document}